\documentclass[letter,nofootinbib,12pt,superscriptaddress]{revtex4-2}

\usepackage[colorlinks,citecolor=blue]{hyperref}
\usepackage{amsmath}
\usepackage{mathrsfs}
\usepackage{bbm}
\usepackage{amsfonts}
\usepackage{amssymb}
\usepackage{latexsym}
\usepackage{graphicx}
\usepackage[english]{babel}
\usepackage{multirow}
\usepackage{float}
\usepackage{url}
\usepackage{slashed}
\usepackage[compat=1.1.0]{tikz-feynman}
\usepackage{orcidlink}

\def\be{\begin{equation}}
\def\ee{\end{equation}}
\def\bea{\begin{eqnarray}}
\def\eea{\end{eqnarray}}

\def\mL{\mathcal{L}}
\def\mH{\mathcal{H}}

\def\mC{\mathcal{C}}
\def\mN{\mathcal{N}}

\def\mM{\mathcal{M}}

\def\zi{\mathrm{i}}
\def\ze{\mathrm{e}}

\def\zd{\mathrm{d}}

\def\zs{\mathrm{s}}

\def\zx{\mathrm{x}}
\def\zy{\mathrm{y}}

\def\zA{\mathrm{A}}
\def\zB{\mathrm{B}}

\def\mbQ{\mathbb{Q}}

\begin{document}

\title{Gravitational wave in symmetric teleparallel gravity with different connections}

\author{Haomin Rao}
\email{rhm137@mail.ustc.edu.cn}
\affiliation{School of Fundamental Physics and Mathematical Sciences,
Hangzhou Institute for Advanced Study, UCAS, Hangzhou 310024, China}
\affiliation{University of Chinese Academy of Sciences, 100190 Beijing, China}

\author{Chunhui Liu}
\email{liuchunhui22@mails.ucas.ac.cn}
\affiliation{School of Fundamental Physics and Mathematical Sciences,
Hangzhou Institute for Advanced Study, UCAS, Hangzhou 310024, China}
\affiliation{University of Chinese Academy of Sciences, 100190 Beijing, China}
\affiliation{Institute of Theoretical Physics, Chinese Academy of Sciences, Beijing 100190, China}

\author{Chao-Qiang Geng}
\email{cqgeng@ucas.ac.cn}
\affiliation{School of Fundamental Physics and Mathematical Sciences,
Hangzhou Institute for Advanced Study, UCAS, Hangzhou 310024, China}

\begin{abstract}
{\color{white} nothing}

We investigate the cosmological perturbations around all three branches of spatially flat universe with different connections
in symmetric teleparallel gravity. The model we consider can cover both the case of $f(\mathbb{Q})$ model
and that of the non-minimal coupling between a scalar field and the non-metricity scalar.
We focus on analyzing and comparing the propagation behavior and stability of the tensorial and non-tensorial gravitational waves on spatially flat universe with different connections.
\end{abstract}

\maketitle

\newpage

\section{Introduction}\label{Introduction}

As the most successful gravity theory, General Relativity (GR) has faced challenges
such as dark matter, dark energy, non-renormalizable problem
and so on \cite{Rubin:1970zza,Planck:2018vyg,SupernovaCosmologyProject:1998vns,Goroff:1985th}.
This has prompted people to explore various modified gravity models.
Within the Riemannian geometric framework on which GR is based,
various modified gravity theories such as the $f(R)$ model
\cite{Sotiriou:2008rp,DeFelice:2010aj,Nojiri:2010wj,Nojiri:2017ncd}
and the scalar-tensor model \cite{Brans:1961sx,Bartolo:1999sq,Kobayashi:2019hrl} have emerged,
and their cosmological applications have also been extensively studied
\cite{Bamba:2010iy,Yang:2010xq,Bamba:2010zxj,Nojiri:2003ft,Nojiri:2005jg,Cognola:2007zu}.
However, Riemannian geometry is not the only geometry that can be used to describe gravity.
Modified gravity theories based on non-Riemannian geometries and their cosmological impact have also received considerable attention
in recent decades \cite{Clifton:2011jh,Bamba:2010wb,Geng:2011aj,Geng:2011ka}.

Symmetric teleparallel gravity (STG) is one of the competing modified gravity schemes,
which identify gravity as non-metricity rather than curvature or torsion.
The simplest STG model with the highest symmetry is equivalent to the GR at level of the equations of motion \cite{Nester:1998mp,BeltranJimenez:2019esp,Capozziello:2022zzh},
and is also called the symmetric teleparallel equivalent of general relativity (STEGR) model.
This fact provides us with another way to modify GR, which is to modify the STEGR model.
Inspired by $f(R)$ and $f(\mathbb{T})$ models,
the most interesting modified STG model is the $f(\mathbb{Q})$ model \cite{BeltranJimenez:2017tkd,Heisenberg:2023lru},
and its cosmological applications have been extensively examined \cite{BeltranJimenez:2019tme,Atayde:2021pgb,Oliveros:2023mwl,Khyllep:2022spx,Zhao:2021zab,Hu:2023ndc}.
Another modified STG model that has also attracted much attention is the non-minimal coupled STG model,
which introduces the non-minimal coupling between a scalar field and the non-metricity scalar,
and its cosmological applications have also been studied extensively \cite{Jarv:2018bgs,Runkla:2018xrv}.
Other more complex and diverse modified STG models can be found in Refs.~\cite{Bahamonde:2022cmz,Armaleo:2023rhj,Li:2021mdp,Li:2022vtn,Chen:2022wtz}.

In most of studies of modified STG models, it has been assumed that the connection $\Gamma^{\rho}_{~\mu\nu}$ is trivially equal to zero in flat universe.
However, even after requiring the connection to be spatially homogeneous and isotropic,
the trivial connection $\Gamma^{\rho}_{~\mu\nu}=0$ is not the only solution in flat universe.
In fact, there are three branch solutions for connections that satisfy spatially homogeneous and isotropic in flat universe \cite{Hohmann:2021ast,Heisenberg:2022mbo,Hohmann:2020zre,Gomes:2023hyk}.
In general, the background evolution of flat universe with different connections is different.
This difference has been preliminary explored in the $f(\mbQ)$ model \cite{Dimakis:2022rkd,Shi:2023kvu} and the non-minimal coupling STG model \cite{Jarv:2023sbp}.

On the other hand, the development of gravitational wave (GW) experiments \cite{LIGOScientific:2016aoc,LIGOScientific:2017vwq,NANOGrav:2023gor,NANOGrav:2023hde}
opens a new window for testing modified gravity theories and cosmological models.
It is foreseeable that within the STG framework,
equipping flat universe with different connections will not only affect the background evolution, but also the propagation behavior of GW.
In this paper, we will investigate cosmological perturbations around all three branches of spatially flat universe with different connections.
The model we consider in this paper can cover both $f(\mathbb{Q})$  and non-minimal coupling STG models.
We will analyse the propagation behavior and stability of the tensorial and non-tensorial gravitational waves on all three branch flat universe.

The present paper is organized as follows.
In section \ref{STGmodel}, after a brief review of symmetric teleparallel gravity,
we introduce the specific model to be considered in this paper.
In order to facilitate the analysis of the propagation behavior of GW in flat universe,
we study the flat universe background of the model in section \ref{Background}.
We will examine all three branches of the spatially flat universe background with different connections.
With the above preparation, in section \ref{Perturbation},
we investigate linear perturbations around all three branches in flat universe.
We force on scalar, vector and tensor perturbations and analyze their propagation behavior and stability.
The conclusion is presented in section \ref{Conclusion}.

In this paper, we adopt the unit $8\pi G=1$ and the signature $(-,+,+,+)$.
The spacetime indices are denoted by Greek indices $\mu, \nu, \rho,...=0, 1, 2, 3$ and the spatial indices are represented by Latin indices $i, j, k,...=1, 2, 3$.
In addition, we distinguish the spacetime affine connection $\Gamma^{\rho}_{~\mu\nu}$
and its associated covariant derivative $\nabla$ from the Levi-Civita connection $\mathring{\Gamma}^{\rho}_{~\mu\nu}$
and its associated covariant derivative $\mathring{\nabla}$, respectively.

\section{Symmetric teleparallel gravity}\label{STGmodel}
The STG theory is formulated in a spacetime endowed with a metric $g_{\mu\nu}$ and an affine connection $\Gamma^{\rho}_{~\mu\nu}$,
which is curvature-free and torsion-free
\be\label{STGconstrain}
R_{\mu\nu\rho}{}^{\sigma}=2\partial_{[\nu}\Gamma^{\sigma}{}_{\mu]\rho}
+2\Gamma^{\sigma}{}_{[\nu|\lambda|}\Gamma^{\lambda}{}_{\mu]\rho}=0,~~
T^{\rho}_{~\mu\nu}=2\Gamma^{\rho}{}_{[\mu\nu]}=0.
\ee
Without curvature and torsion, the gravity is identified with non-metircity
\be\label{nonmetricity}
Q_{\rho\mu\nu}=\nabla_{\rho}g_{\mu\nu}.
\ee
The STG constraints in Eq.~(\ref{STGconstrain}) indicate that the affine connection $\Gamma^{\rho}_{~\mu\nu}$ can be in general expressed as
\be\label{STGconnection}
\Gamma^{\rho}_{~\mu\nu}=\frac{\partial x^{\rho}}{\partial y^{\sigma}}\partial_{\mu}\partial_{\nu}y^{\sigma},
\ee
where $\{ y^{\mu}(x)\}$  can be regarded as a coordinate system.
In the coordinate system $\{ y^{\mu}\}$, the connection is trivial, that is, $\Gamma^{\rho}_{~\mu\nu}=0$, which is also called the coincident gauge.
The metric $g_{\mu\nu}$ and functions $y^{\mu}$ can be regarded as basic variables in the STG theory.

The simplest STG model is the so-called symmetric teleparallel equivalent of general relativity (STEGR) model, whose action is
\be\label{STEGR}
S_{\text{STEGR}}=\frac{1}{2}\int \zd^4x \sqrt{-g}\, \mathbb{Q}+S_{m},
\ee
where the non-metricity scalar $\mathbb{Q}$ is defined as
\be\label{nonmetricity}
\mathbb{Q}=P^{\rho\mu\nu}Q_{\rho\mu\nu}
=-\frac{1}{4}Q_{\rho\mu\nu}Q^{\rho\mu\nu}+\frac{1}{2}Q_{\rho\mu\nu}Q^{\mu\nu\rho}+\frac{1}{4}Q_{\mu}Q^{\mu}-\frac{1}{2}Q_{\mu}\tilde{Q}^{\mu},
\ee
with the non-metricity conjugate $P^{\rho\mu\nu}=-(1/4)Q^{\rho\mu\nu}+(1/2)Q^{(\mu\nu)\rho}+(1/4)(Q^{\rho}-\tilde{Q}^{\rho})g^{\mu\nu}-(1/4)g^{\rho(\mu}Q^{\nu)}$,
$Q_{\mu}=Q_{\mu\nu}{}^{\nu}$, $\tilde{Q}_{\mu}=Q^{\nu}{}_{\nu\mu}$,
and other matter with the action $S_m$ assumed to be minimally coupled to the metric.
Since we have the identity
\be
\mathring{R}\equiv\mathbb{Q}-\mathring{\nabla}_{\mu}(Q^{\mu}-\tilde{Q}^{\mu}),
\ee
the action (\ref{STEGR}) is identical to the Einstein-Hilbert action up to a surface term,
where the curvature scalar $\mathring{R}$ is defined by the Levi-Civita connection and considered as being fully constructed from the metric.
Since the surface term in the action does not affect the equations of motion,
we say that  STEGR is equivalent to GR at the level of equations of motion.

The coincidence that the STEGR model is equivalent to GR provides another way to modify GR,
which is to modify the STEGR model within the STG framework.
Along this line, a variety of modified STG models have been proposed.
The most interesting one is the $f(\mathbb{Q})$ model, which generalizes $\mathbb{Q}$ in the action (\ref{STEGR}) to a smooth function $f(\mathbb{Q})$.
Another widely studied scheme to modify STEGR is to introduce the non-minimal coupling between a scalar field and the non-metricity scalar.

In this paper, we concentrate on the following modified STG model
\be\label{model}
S=\int \zd^4x \sqrt{-g}\,\left[ \frac{1}{2}f(\mathbb{Q})+\frac{1}{2}\mC(\phi)\mathbb{Q}+\mL(X,\phi)\right]+S_m,
\ee
where $X=-g^{\mu\nu}\partial_{\mu}\phi\partial_{\nu}\phi$,
$f(\mathbb{Q})$ and $\mC(\phi)$ are the smooth functions of $\mathbb{Q}$ and the scalar field $\phi$,  respectively.
Both $f(\mathbb{Q})$ and non-minimal coupling STG models can be regarded as the submodel of the model (\ref{model}).
This makes the analysis of cosmological pertuabtions below available to both $f(\mathbb{Q})$  and non-minimal coupling STG models.

Equations of motion of the model (\ref{model}) can be obtained from the variations with respect to
basic variables $g_{\mu\nu}$, $y^{\mu}$ and $\phi$.
The equation of motion follow from the variations with respect to metric $g_{\mu\nu}$ is
\be\label{Eom1}
(f_{\mathbb{Q}}+\mC)\mathring{G}^{\mu\nu}+\frac{1}{2}(\mathbb{Q}f_{\mathbb{Q}}-f)g^{\mu\nu}
+2\nabla_{\rho}(f_{\mathbb{Q}}+\mC)P^{\rho\mu\nu}=\Theta_{\phi}^{\mu\nu}+\Theta_{m}^{\mu\nu},
\ee
where
$f_{\mathbb{Q}}=\zd f(\mathbb{Q})/\zd \mathbb{Q}$,  
$\mathring{G}^{\mu\nu}=\mathring{R}^{\mu\nu}-(1/2)\mathring{R}g^{\mu\nu}$ is the Einstein tensor,
$\Theta_{m}^{\mu\nu}=(2/\sqrt{-g})(\delta S_m/\delta g_{\mu\nu})$ is the energy-momentum tensor for other matters,
$\Theta_{\phi}^{\mu\nu}=2\mL_{X}\partial^{\mu}\phi\partial^{\nu}\phi+\mL g^{\mu\nu}$ is the energy-momentum tensor for the scalar field $\phi$,
where $\mL_{X}=\partial{\mL}(X,\phi)/\partial X$.
The one follow from the variations with respect to $y^{\mu}$ is
\be\label{Eom2}
\nabla_{\mu}\nabla_{\nu}\left[\sqrt{-g}(f_{\mathbb{Q}}+\mC)P^{\mu\nu}{}_{\rho}\right]=0.
\ee
Here we have assumed that there is no direct coupling between other matter and the connection,
so $S_{m}$ does not play any role in the variations with respect to the connection.
Finally, the equation of motion from the variations with respect to the scalar field $\phi$ is
\be\label{Eom3}
\mathring{\nabla}_{\mu}\left(2\mL_{X}\mathring{\nabla}^{\mu}\phi\right)+\mL_{\phi}+\frac{1}{2} \mathbb{Q}\, \mC_{\phi}=0
\ee
where $\mL_{\phi}=\partial{\mL}(X,\phi)/\partial \phi$ and $\mC_{\phi}=\zd \mC(\phi)/\zd \phi$.
All of these equations of motion are consistent with the covariant conservation law $\mathring{\nabla}_{\mu}\Theta_{m}^{\mu\nu}=0$
and the Bianchi identities $\mathring{\nabla}_{\mu}\mathring{G}^{\mu\nu}\equiv0$
and $\nabla_{\mu}\nabla_{\nu}(\sqrt{-g}P^{\mu\nu}{}_{\rho})\equiv0$.
Vice versa, under the premise that the covariant conservation law $\mathring{\nabla}_{\mu}\Theta_{m}^{\mu\nu}=0$ is satisfied,
Eq.~(\ref{Eom2}) can be derived from Eqs.~(\ref{Eom1}) and (\ref{Eom3}).

\section{Cosmological background}\label{Background}

Now we apply the model (\ref{model}) to flat universe. In this section, we
only consider the background evolutions and leave the discussions on the cosmological perturbations to next sections.
The background equations obtained in this section will be a great help in simplifying the quadratic action of perturbations in the next section.

In flat universe, the metric can be expressed in rectangular coordinate as
\be\label{FRWmetric}
\zd s^{2}=g_{\mu\nu}\zd x^{\mu}\zd x^{\nu}=a^{2}\left(-\zd\eta^{2}+\delta_{ij} \zd x^{i} \zd x^{j}\right),
\ee
where $a=a(\eta)$ is the scale factor and $\eta$ is the conformal time.
There are 6 Killing vector fields $\{\xi_{I}^{\mu}, I=1,2...6\}$ in flat universe, which can be expressed as
\be\label{killingvector}
\xi^{\mu}_{I}=\delta^{\, \mu}_{~I}~,~
\xi^{\mu}_{I+3}=\epsilon_{Iij}\delta^{\mu}_{~i}x^{j}~,~~~~ I=1,2,3
\ee
It can be proved that the metric (\ref{FRWmetric}) satisfies the condition: $\mathcal{L}_{\xi_{I}}g_{\mu\nu}=0$,
where $\mathcal{L}_{\xi_{I}}$ is the Lie derivative along the Killing vector field $\xi^{\mu}_{I}$.
This reflects the fact that the metric is homogeneous and isotropic.

Unlike the case of Riemannian geometry, in the STG theory, the affine connection is still arbitrary to some extent after the metric is determined.
As suggested in Refs.~\cite{Hohmann:2021ast,Heisenberg:2022mbo,Hohmann:2020zre,Gomes:2023hyk}, it is natural to further require that the connection is also homogeneous and isotropic, that is,
\be\label{FRWconnection0}
\mathcal{L}_{\xi_I}{\Gamma}^{\rho}_{~\mu\nu}={\nabla}_{\mu}{\nabla}_{\nu}\,\xi^{\rho}_{I}
-\xi^{\sigma}_{I}R_{\sigma\mu\nu}{}^{\rho}-\nabla_{\mu}(T^{\rho}_{~\nu\sigma}\xi^{\sigma}_{I})
=0~,~~~~ I=1,2...6.
\ee
Although ${\Gamma}^{\rho}_{~\mu\nu}$ is coordinate dependent, the Lie derivative of ${\Gamma}^{\rho}_{~\mu\nu}$ does not depend on the coordinate.
Hence, the condition (\ref{FRWconnection0}) is unambiguous.
Combining the condition (\ref{FRWconnection0}) and the STG constraints (\ref{STGconstrain}),
the non-zero component of the connection ${\Gamma}^{\rho}_{~\mu\nu}$ can be obtained as
\be\label{FRWconnection}
\Gamma^{0}_{~00}=K_{1}(\eta)~,~\Gamma^{0}_{~ij}=K_{2}(\eta)\delta_{ij}~,~\Gamma^{i}_{~0j}=\Gamma^{i}_{~j0}=K_{3}(\eta)\delta_{ij},
\ee
where $\{K_1(\eta), K_2(\eta), K_3(\eta)\}$ have three branch solutions, given by
\bea
& &
\text{branch 1 :}~ K_1=\gamma~,~K_2=0~,~K_3=0,
\\ & &
\text{branch 2 :}~ K_1=\gamma'/\gamma+\gamma~,~K_2=0~,~K_3=\gamma,
\\ & &
\text{branch 3 :}~ K_1=-\gamma'/\gamma~,~K_2=\gamma~,~K_3=0,
\eea
where $\gamma=\gamma(\eta)$ is a function of the conformal time $\eta$
and the superscript ``prime" represents the derivative with respect to the conformal time $\eta$.

Putting the metric (\ref{FRWmetric}) and the solution (\ref{FRWconnection}) into Eqs.~(\ref{Eom1})-(\ref{Eom3}), we obtain the background equations as
\bea
& &\label{beom1}
3(f_{\mathbb{Q}}+\mC)\mH^{2}+\frac{1}{2}a^{2}(f-\mathbb{Q}f_{\mbQ})+\frac{3}{2}(K_{3}-K_{2})(f'_{\mbQ}+\mC')=a^{2}(\rho_{\phi}+\rho_{m}),
\\ & &\label{beom2}
(f_{\mathbb{Q}}+\mC)(2\mH'+\mH^{2})+\frac{1}{2}a^{2}(f-\mbQ f_{\mbQ})+\frac{1}{2}(4\mH-K_{2}-3K_{3})(f'_{\mbQ}+\mC')=-a^{2}(p_{\phi}+p_{m}),~~~~~
\\ & &\label{beom3}
(K_{2}-K_{3})(f_{\mbQ}''+\mC'')+2\left[\mH(K_{2}-K_{3})-K_{1}K_{2}\right](f_{\mbQ}'+\mC')=0,
\\ & &\label{beom4}
2\mL_{X}\phi''+2(\mL_{X}'+2\mH \mL_{X})\phi'-a^{2}\mL_{\phi}-\frac{1}{2}a^{2}\mathbb{Q}\, \mC_{\phi}=0,
\eea
where
\be\label{Qs0}
\mbQ=-\frac{3}{a^{2}}\left[2\mH^{2}-2\mH(K_{2}+K_{3})+K_{1}K_{2}-K_{1}K_{3}+K_{3}^{2}\right],
\ee
and $\mathcal{H}=a'/a$ is the conformal Hubble rate, $\rho_{\phi}=2X\mL_{X}-\mL$ is the energy density of the scalar field $\phi$,
$p_{\phi}=\mL$ is the pressure of the scalar field $\phi$,
$\rho_{m}$ and $p_{m}$ are the energy density and pressure of other matter, respectively.
Note that in Eq.~(\ref{Qs0}), we have used the condition $K_{2}K_{3}=0$ that is satisfied by all three branch solutions.

Given the equations of state of other matter and appropriate initial conditions, Eqs.~(\ref{beom1})-(\ref{beom4}) can tell us how
$a$, $\gamma$ and $\phi$ evolve over time.
The specific evolution behavior will depend on the specific form of $f(\mbQ)$ and $\mC(\phi)$ and which branch of connection we take.
In particular, for the branch 1, the function $\gamma$ does not appear in Eqs.~(\ref{beom1})-(\ref{beom4}) and the non-metricity scalar $\mbQ$.
Therefore, for the branch 1, the function $\gamma$ has no effect on the background evolution,
and the background equations are exactly the same as the trivial case of $\Gamma^{\rho}_{~\mu\nu}=0$.
This is the most studied case in previous studies of STG cosmology.
But for the branch 2 and branch 3, $\gamma$ is determined by the equations of motion,
and the background evolution behavior will be different from the case of $\Gamma^{\rho}_{~\mu\nu}=0$.
More specific research on background evolutions of branch 2 and branch 3 can be found in Refs.~\cite{Dimakis:2022rkd,Shi:2023kvu,Jarv:2023sbp}.

\section{Cosmological perturbations}\label{Perturbation}

In order to analyze the propagation behavior and stability of GW on different branches,
we investigate the cosmological perturbations of the model (\ref{model}) around flat universe with different connection branches in this section.
For simplicity, we ignore other matter so that $S_m=0$.
We also split the total action into $S=S_{\mbQ}+S_{\phi}$,
where $S_{\mbQ}$ is the part that contains $\mbQ$, and $S_{\phi}$ is the part that only contain the scalar field $\phi$.

We can parameterize the metric after considering perturbations as
\bea\label{metricperturbation}
& &g_{00}=-a^{2}(1+2A),~ g_{0i}=a^{2}(B_{,i}+B_{i}^{V}),\nonumber\\
& &g_{ij}=a^{2}\left[(1-2\psi)\delta_{ij}+2E_{,ij}+E_{i,j}^{V}+E_{j,i}^{V}+h^{T}_{ij}\right],
\eea
the subscript $``,i"$ means $\partial_{i}$.
All the vector perturbations are transverse and denoted by the superscript $V$, both the
tensor perturbations are transverse and traceless and denoted by the superscript $T$.
In addition, the scalar field $\phi$ is decomposed as $\phi=\bar{\phi}+\delta\phi$.
In principle, we should also consider the perturbation of connection $\Gamma^{\rho}_{~\mu\nu}=\bar{\Gamma}^{\rho}_{~\mu\nu}+\delta\Gamma^{\rho}_{~\mu\nu}$.
But due to diffeomorphism symmetry, we can always choose the gauge $\delta\Gamma^{\rho}_{~\mu\nu}=0$.
Note that such a choice has exhausted all diffeomorphism symmetries,
so all metric perturbations in Eq.~(\ref{metricperturbation}) can no longer be eliminated by gauge transformations.

\subsection{Tensor perturbation}

For tensor perturbations, we can expand them as follows
\be\label{Texpand}
h^{T}_{ij}(\eta, \vec{x})=\sum_{\zA=+,\times}\int \frac{\zd^{3}k}{(2\pi)^{\frac{3}{2}}}\, h_{\zA}(\eta, \vec{k})\, \hat{e}_{i j}^{\zA}(\vec{k})\,
\ze^{\zi\vec{k}\cdot\vec{x}},
\ee
where the tensor polarization bases $\{\hat{e}_{ij}^{\zA}(\vec{k}), \zA=+,\times\}$ satisfy the relation
$\hat{e}_{ij}^{\zA}(\vec{k})\hat{e}_{ij}^{\zB}(\vec{k})=2\delta^{\zA\zB}$.
Then, the quadratic action of $S_{\mbQ}$ for tensor perturbations can be directly obtained as
\bea\label{SQt}
& &
S_{\mbQ,T}^{(2)}=\sum_{\zA=+,\times} \int \zd\eta\, \zd^{3}k\,
\frac{a^{2}}{4}(f_{\mbQ}+\mC)\left(\left|h_{\zA}'\right|^{2}-{\omega}^{2}_{\mbQ}\left|h_{\zA}\right|^{2}\right),
\\ & & \nonumber
\text{with}~~  {\omega}^{2}_{\mbQ}=k^{2}+2(2\mH'+\mH^{2})+(4\mH+K_{2}-3K_{3})[\ln(f_{\mbQ}+\mC)]'+a^{2}(f-\mbQ f_{\mbQ})/(f_{\mbQ}+\mC),
\eea
while the quadratic action of $S_{\phi}$ for tensor perturbations is given by
\be\label{Sphit}
S_{\phi,T}^{(2)}=-\sum_{\zA=+,\times} \int \zd\eta\, \zd^{3}k\,
\frac{a^{4}}{2}p_{\phi}\left|h_{\zA}\right|^{2}
\ee
Using the background equation (\ref{beom2}), the total quadratic action for tensor perturbations can be simplified as
\be\label{ST}
S_{T}^{(2)}=\sum_{\zA=+,\times} \int \zd\eta\, \zd^{3}k\,
\frac{a^{2}}{4}(f_{\mbQ}+\mC)\left(\left|h_{\zA}'\right|^{2}-{\omega}^{2}\left|h_{\zA}\right|^{2}\right),
~~\text{with}~~{\omega}^{2}=k^{2}+2K_{2}[\ln(f_{\mbQ}+\mC)]'.
\ee
The propagation equation of GW is obtained from the action (\ref{ST}) as
\be\label{tensoreom}
h_{\zA}''+\left(2\mH+[\ln(f_{\mbQ}+\mC)]'\right)h_{\zA}'+\omega^{2}h_{\zA}=0.
\ee
When $f(\mbQ)\propto\mbQ$ and $\mC(\phi)=\text{constant}$, Eq.~(\ref{tensoreom}) can be reduced to the GW propagation equation in GR.
Next we consider the case where $f(\mbQ)$ and $\mC(\phi)$ are more general functions.

Compared with the GW propagation equation in GR, Eq.~(\ref{tensoreom}) shows that
the STG model (\ref{model}) brings two types of corrections to the GW propagation.
The first corrections is that the STG model (\ref{model}) modifies the friction term in the GW propagation equation.
The modified friction term will affect the amplitude of GW,
making the GW luminosity distance $d_{L}^{GW}$ no longer equal to the standard electromagnetic luminosity distance $d_{L}^{EM}$
\cite{Belgacem:2017ihm,Belgacem:2018lbp}
\be
d_{L}^{GW}(z)=d_{L}^{EM}(z)\exp\left\{ \int^{z}_{0} \frac{\zd \bar{z}}{1+\bar{z}}\delta(\bar{z})\right\}~,~~\text{with}~~
\delta(z)=\frac{1}{2\mH}[\ln(f_{\mbQ}+\mC)]',
\ee
where $z$ is the redshift.
This correction appears in both $f(\mbQ)$ and non-minimal coupling STG models, and exists in all three branches of the background connection.

The second correction is that the STG model (\ref{model}) modifies the dispersion relation in the GW propagation equation.
It can be seen that there is an additional time-dependent mass term in the dispersion relation of GW in Eq.~(\ref{ST}).
This makes the phase velocity $v_{p}$ and group velocity $v_{g}$ of GW different from the speed of light:
\be\label{vtp}
v_{p}=\frac{\omega}{k}\approx 1+\frac{K_{2}[\ln(f_{\mbQ}+\mC)]'}{k^{2}}~~,~~
v_{g}=\frac{\zd \omega}{\zd k}\approx 1-\frac{K_{2}[\ln(f_{\mbQ}+\mC)]'}{k^{2}}.
\ee
Note that we have assumed $k^{2}\gg|K_{2}[\ln(f_{\mbQ}+\mC)]'|$ in Eq.~(\ref{vtp}).
Unlike the correction to the friction term,
the correction to the dispersion relation only exists in the branch 3,
because $K_{2}$ is non-vanishing only in the branch 3.
In the branch 1 and branch 2 flat universe, the speed of GW is the same as the speed of light.
Such a result holds true for both $f(\mbQ)$ model and  non-minimal coupling STG models.
In addition, the difference between the speed of GW and the speed of light can be tightly constrained by the present GW experiments
\cite{LIGOScientific:2017zic,Zhu:2023wci}.
This would impose additional strong constraints on the evolution of the branch 3 flat universe.

It should be emphasized that all corrections to the propagation of GW depend on the specific background evolution,
that is, the specific expression of $\mH(\eta)$ and $\gamma(\eta)$, etc.
The latter in turn depends on the specific STG model, that is, the specific forms of $f(\mbQ)$ and $\mC(\phi)$.
In addition, no matter which model we adopt, $\gamma(\eta)$ in the branch 1 appears
neither in the background equation nor in the GW propagation equation.
This means that for the branch 1 flat universe,
both the background evolution and the GW propagation behavior are exactly the same as the case of the trivial connection  $\Gamma^{\rho}_{~\mu\nu}=0$.

Finally, let's briefly discuss the stability of tensor perturbations.
It can be seen from the action (\ref{ST}) that we need to require $f_{\mbQ}+\mC>0$ to avoid ghost instability in tensor perturbations.
In addition, for the case of $K_{2}[\ln(f_{\mbQ}+\mC)]'\lesssim -\mH$,
 tensor perturbations have the risk of tachyon instability.

\subsection{Vector perturbation}
For vector perturbations, we can expand them with the vector polarization bases, such as
\be\label{Vexpand}
B_{i}^{V}(\eta, \vec{x})=\sum_{\zA=\zx,\zy}\int \frac{\zd^{3}k}{(2\pi)^{\frac{3}{2}}}\,
B_{\zA}(\eta, \vec{k})\, \hat{e}_{i}^{\zA}(\vec{k})\,\ze^{\zi\vec{k}\cdot\vec{x}},
\ee
where the vector polarization bases $\{\hat{e}_{i}^{\zA}(\vec{k}), \zA=\zx,\zy\}$ satisfy the relation
$\hat{e}_{i}^{\zA}(\vec{k})\hat{e}_{i}^{\zB}(\vec{k})=\delta^{\zA\zB}$.
Then, the quadratic action of $S_{\mbQ}$ for vector perturbations can be directly obtained as
\bea\label{SQv}
& &
S_{\mbQ,V}^{(2)}=\sum_{\zA=\zx,\zy} \int \zd\eta\, \zd^{3}k\,
a^{2}\left\{\frac{1}{4}(f_{\mbQ}+\mC)k^{2}\left|B_{\zA}-E_{\zA}'\right|^{2}+\frac{1}{2}m_{B}^{2}|B_{\zA}|^{2}-\frac{1}{2}k^{2}m_{E}^{2}|E_{\zA}|^{2}\right\},
\\ & & \nonumber
\text{with}~~~ m_{B}^{2}=3(f_{\mathbb{Q}}+\mC)\mH^{2}+\frac{1}{2}a^{2}(f-\mathbb{Q}f_{\mbQ})+\frac{1}{2}(3K_{3}-K_{2})(f'_{\mbQ}+\mC'),
\\ & & \nonumber
~~~~~~~~~ m_{E}^{2}=(f_{\mathbb{Q}}+\mC)(2\mH'+\mH^{2})+\frac{1}{2}a^{2}(f-\mbQ f_{\mbQ})+\frac{1}{2}(4\mH+K_{2}-3K_{3})(f'_{\mbQ}+\mC').
\eea
And the quadratic action of $S_{\phi}$ for vector perturbations can also be directly found to be
\be\label{Sphiv}
S_{\phi,V}^{(2)}=-\sum_{\zA=\zx,\zy} \int \zd\eta\, \zd^{3}k\,
a^{4}\left\{\frac{1}{2}\rho_{\phi}|B_{\zA}|^{2}+\frac{1}{2}k^{2}p_{\phi}|E_{\zA}|^{2}  \right\}.
\ee
Using the background equations (\ref{beom1}) and (\ref{beom2}), the total quadratic action for vector perturbations is simplified as
\be\label{SV1}
S_{V}^{(2)}=\sum_{\zA=\zx,\zy} \int \zd\eta\, \zd^{3}k\,
a^{2}\left\{\frac{1}{4}(f_{\mbQ}+\mC)k^{2}\left|B_{\zA}-E_{\zA}'\right|^{2}+\frac{K_{2}}{2}(f_{\mbQ}'+\mC')(|B_{\zA}|^{2}-k^{2}|E_{\zA}|^{2})\right\}.~~
\ee
It can be seen that $B_{\zA}$ is non-dynamic field and the variation of the action (\ref{SV1}) with respect to $B_{\zA}$ leads to the following constraint
\be\label{Vceq1}
B_{\zA}=\frac{E_{\zA}'}{1+2k^{-2}K_{2}[\ln(f_{\mbQ}+\mC)]'}
\ee
One can eliminate $B_{\zA}$ from the action (\ref{SV1}) by substituting the constraint (\ref{Vceq1}) back into it.
After that, the quadratic action for vector perturbations can be simplified as
\be\label{Vceq2}
S_{V}^{(2)}=\sum_{\zA=\zx,\zy} \int \zd\eta\, \zd^{3}k\,
\frac{1}{2}\tilde{a}^{2}\left(\left|E_{\zA}'\right|^{2}-\omega^{2}\left|E_{\zA}\right|^{2}\right),
\ee
where
\be\nonumber
\tilde{a}^{2}=\frac{a^{2}K_{2}(f_{\mbQ}+\mC)'}{1+2k^{-2}K_{2}[\ln(f_{\mbQ}+\mC)]'}~~,~~{\omega}^{2}=k^{2}+2K_{2}[\ln(f_{\mbQ}+\mC)]'.
\ee
When $f(\mbQ)\propto\mbQ$ and $\mC(\phi)=\text{constant}$, Eq.~(\ref{tensoreom}) is reduced to $S^{(2)}_{V}=0$, that is,
there is no dynamic degree of freedom in vector perturbations, and this is exactly the case of GR.
Next we consider the case where $f(\mbQ)$ and $\mC(\phi)$ are more general functions.

In the branch 3 flat universe, usually $K_{2}=\gamma\neq0$,  so there are two dynamic degrees of freedom in vector perturbations.
These two vector degrees of freedom can be regarded as vectorial GW, and there are already some experiments to detect and constrain them
\cite{Chen:2021wdo,Wu:2021kmd,Chen:2023uiz}.
To avoid ghost instability, we need to require $\tilde{a}^{2}>0$, which  means $K_{2}(f_{\mbQ}+\mC))'>0$ for large $k$ modes.
It is interesting to note that the vectorial and tensorial GW have the same dispersion relation $\omega^{2}$.
This means that the vectorial and tensorial GW have the same propagation speed and are different from the speed of light.

In the branch 1 and branch 2 flat universe, $K_{2}=0 \Rightarrow S^{(2)}_{V}=0$, so there is no dynamic degree of freedom in vector perturbations.
This lack of degrees of freedom implies that vector perturbations suffer from the strong coupling problem in the branch 1 and branch 2 flat universe.

\subsection{Scalar perturbations}\label{Sperturbation}

Finally, we examine the stability of scalar perturbations.
Without loss of generality, we take the Lagrangian of the scalar field in action (\ref{model}) as $\mL(X,\phi)=\frac{1}{2}X-V(\phi)$.
Since the quadratic action for scalar perturbations is very complicated and bloated, we only show the key parts of the quadratic action in this section.

For scalar perturbations, the quadratic action can be directly obtained as
\bea\label{SS1}
& &\nonumber
S_{V}^{(2)}=\frac{1}{2} \int \zd\eta\, \zd^{3}k\,
\bigg[ \frac{9}{2}f_{\mbQ\mbQ}(K_{2}^{2}+K_{3}^{2}){A'}^{2}+f_{\mbQ\mbQ}\bar{\mN}_{1}A'(3\psi'+k^{2}E')+k^{2}f_{\mbQ\mbQ}\bar{\mN}_{2}A'B
\\ & &\quad\quad\quad\quad\quad\quad\quad
+4k^{2}a^{2}(\mC+f_{\mbQ})\psi'B-k^{2}f_{\mbQ\mbQ}\bar{\mN}_{3}(\psi'+k^{2}E')B+k^{2}\bar{\mN}_{4}B^{2}+\cdot\cdot\cdot\bigg],
\eea
where
\bea
& &\nonumber
\bar{\mN}_{1}=3\left[3K_{3}^{2}-K_{2}^{2}+4\mH(K_{2}-K_{3})\right],
\\& &\nonumber
\bar{\mN}_{2}=3\left[K_{1}K_{2}-K_{1}K_{3}-K_{3}^{2}+2\mH(K_{3}-K_{2})\right],
\\& &\nonumber
\bar{\mN}_{3}=3K_{3}^{2}+K_{1}K_{2}+3K_{1}K_{3}-2\mH(2K_{1}+K_{2}+5K_{3})+8\mH^{2},
\\& &\nonumber
\bar{\mN}_{4}=(1/2)k^{2}f_{\mbQ\mbQ}(K_{1}+K_{3}-2\mH)^{2}+a^{2}K_{2}(\mC+f_{\mbQ})'.
\eea
In the above expression, we only show the terms containing $A'$, $B\psi'$, $BE'$ and $B^{2}$, while other terms are represented by ``$\cdot\cdot\cdot$".
We also abbreviate $A^{*}B$ as $AB$, $B^{*}B$ as $B^{2}$, and so on.
From the action (\ref{SS1}), we know that the value of $f_{\mbQ\mbQ}$ determines whether the variable $A$ is dynamic,
so we will discuss it case by case below.

\subsubsection{The case of $f_{\mbQ\mbQ}\neq0$}\label{Scase1}

For the case of $f_{\mbQ\mbQ}\neq0$, only $B$ is non-dynamic among all scalar perturbations.
The variations of the action (\ref{SS1}) with $B$ lead to the following constraint
\be\label{Sceq1}
B=-\frac{1}{2\bar{\mN}_{4}}\left[f_{\mbQ\mbQ}\bar{\mN}_{2}A'+4a^{2}(f_{\mbQ}+\mC)\psi'-f_{\mbQ\mbQ}\bar{\mN}_{3}(3\psi'+k^{2}E')\right]+\cdot\cdot\cdot.
\ee
The ellipses in Eq.~(\ref{Sceq1}) indicate terms that do not contain the time derivatives of perturbation variables.
Since only the kinetic energy term is important in the following,
we only show the terms that will contribute to the kinetic energy here.
After substituting the constraints (\ref{Sceq1}) back into the action (\ref{SS1}), the quadratic action for scalar perturbations can be expressed as
\be\label{SS2}
S_{S}^{(2)}=\frac{1}{2}\sum_{\zs_{1}=1}^{4}\sum_{\zs_{1}=1}^{4}\int \zd\eta\, \zd^{3}k~
a^{2}\left( \mM_{\zs_{1}\zs_{2}}\Phi_{\zs_{1}}'\Phi_{\zs_{2}}'+\cdot\cdot\cdot \right),~~~
\text{where}~~\Phi_{\zs}=(\delta\phi, A, \psi, E).
\ee
Here, we only show the kinetic energy terms.
Using the background equations (\ref{beom1})-(\ref{beom4}), the determinant of the kinetic energy matrix $\mM$ can be simplified to
\be\label{det1}
\det\mM =- \frac{36k^{4}K_{2}^{2}(f_{\mbQ}+\mC)^{2}f_{\mbQ\mbQ}(f_{\mbQ}+\mC)'}{2a^{2}K_{2}(f_{\mbQ}+\mC)'+k^{2}f_{\mbQ\mbQ}(2\mH-K_{1})^{2}}.
\ee

In the branch 3 flat universe, usually $K_{2}\neq0 \Rightarrow \det\mM\neq0 $, so there are four dynamical degrees of freedom in scalar perturbations.
One degree of freedom comes from the scalar field $\phi$, and the other three come from the metric.
For large $k$ modes, the determinant of the kinetic energy matrix $\mM$ can be further simplified to
\be\label{det2}
\det\mM \overset{\text{UV}}{\approx} -\frac{36k^{2}K_{2}^{3}(f_{\mbQ}+\mC)^{2}(f_{\mbQ}+\mC)'}{(2\mH-K_{1})^{2}}.
\ee
From the analysis in the previous subsections,
we already know that in order to ensure that tensor and vector perturbations are ghost-free,
we need to require $f_{\mbQ}+\mC>0$ and $K_{2}(f_{\mbQ}+\mC)'>0$.
However, these two conditions make $\det\mM<0$ in Eq.~(\ref{det2}).
It means that there is at least one ghost mode in scalar perturbations.
And the ghost mode in scalar perturbations can only arise from the metric,
because the scalar field with the Lagrangian $\mL(X,\phi)=\frac{1}{2}X-V(\phi)$ will not bring the ghost mode.
Therefore, ghost instability is inevitable in the branch 3 flat universe.

In the branch 1 and branch 2 flat universe, $K_{2}=0 \Rightarrow \det\mM=0$, so at most only three dynamic degrees of freedom appear in linear order.
This lack of degrees of freedom implies that scalar perturbations suffer from the strong coupling problem in the branch 1 and branch 2 flat universe.
This is very similar to the case of vector perturbation.

To sum up, the non-trivial (i.e. $f_{\mbQ\mbQ}\neq 0$) model (\ref{model}) will suffer from the strong coupling issue in the branch 1 and branch 2 flat universe,
and will suffer from the problem of ghost instability in the branch 3 universe.
Such a result holds true for the pure $f(\mbQ)$ model, that is, the model with $\mC(\phi)=0$.
The stability of the $f(\mbQ)$ model is also discussed in Ref.~\cite{Gomes:2023tur} at almost the same time, and we reached the same conclusion.
They argue that these instabilities should also exist in more general modified STG models.
Here we show that accounting for the non-minimal coupling indeed does not cure these instabilities.
These instabilities may arise from the fact that
non-trivial STG models always lack diffeomorphism (coincident gauge) or contain higher-order derivatives (via Eq.~(\ref{STGconnection})).

\subsubsection{The case of $f_{\mbQ\mbQ}=0$}\label{Scase2}

For the case of $f_{\mbQ\mbQ}=0$, $f$ is a linear function of $\mbQ$ (constant term can be absorbed into $V(\phi)$).
It means that $f$  can be absorbed into $\mC\mbQ$ in the action (\ref{model}). So we can only consider the non-minimal coupling STG model.

It can be seen from the action (\ref{SS1}) that in this case, both variables $A$ and $B$ are non-dynamic.
The variations of the action (\ref{SS1}) with them lead to the following constraints
\bea
& &\label{Sceq2}
(\mC'+4\mH\mC)A+2K_{2}\mC'B+4\mC\psi'+\cdot\cdot\cdot=0,
\\ & &\label{Sceq3}
2({\phi'}^{2}-6\mH^{2}\mC-3K_{3}\mC')A+k^{2}(\mC'+4\mH\mC)B-4\mH\mC(3\psi'+k^{2}E')-2\phi'\delta\phi'+\cdot\cdot\cdot=0.~~~~~
\eea
The ellipses indicate terms that do not contain the time derivatives of perturbation variables.
These constraint equations are just used to solve the non-dynamic variables $A$ and $B$.
One can eliminate these two non-dynamic variables from the action (\ref{SS1}) by substituting
the constraints (\ref{Sceq2}) and (\ref{Sceq3}) back into it.
After that, the quadratic action for scalar perturbations can be expressed as
\be\label{SS3}
S_{S}^{(2)}=\frac{1}{2}\sum_{\zs_{1}=1}^{3}\sum_{\zs_{1}=1}^{3}\int \zd\eta\, \zd^{3}k~
a^{2}\left( \tilde{\mM}_{\zs_{1}\zs_{2}}\tilde{\Phi}_{\zs_{1}}'\tilde{\Phi}_{\zs_{2}}'+\cdot\cdot\cdot \right),~~~
\text{where}~~\tilde{\Phi}_{\zs}=(\delta\phi, \psi, E).
\ee
Here, we only show the kinetic energy terms.
No matter in which branch, for large $k$ modes, the determinant of the kinetic energy matrix $\tilde{\mM}$ can be approximately written as
\be\label{det3}
\det\tilde{\mM} \overset{\text{UV}}{\approx}-\frac{4k^{4}\mC^{2}{\mC'}^{2}}{(\mC'+4\mH\mC)^{2}}.
\ee

For the general case where $\mC$ is not a constant, $\det\tilde{\mM}<0$.
It means that there are three dynamic degrees of freedom in scalar perturbations, and at least one of them is a ghost mode.
This ghost mode can only come from the metric, because the scalar field $\phi$ with the Lagrangian $\mL(X,\phi)=\frac{1}{2}X-V(\phi)$ cannot cause the ghost mode.
Therefore, in the non-minimal coupling STG model,
there is always ghost instability in scalar perturbations, no matter which branch flat universe it is in.

For the case where $\mC$ is a constant, the kinetic energy matrix $\tilde{\mM}$ satisfies
\be\label{eigen}
\text{rank of}~\tilde{\mM}=1~~,~~\text{non-zero eigenvalue of}~ \tilde{\mM}=1+\frac{{\phi'}^{2}}{\mH^{2}}>0.
\ee
This is precisely the result in GR. There is only one dynamic degree of freedom brought by the scalar field $\phi$, and this degree of freedom is ghost-free.

In summary, the non-minimally coupled STG model suffers from the problem of ghost instability, unless the model is reduced to GR.

\section{Conclusion}\label{Conclusion}

In this paper, we have considered a STG model that can encompass both $f(\mathbb{Q})$ and  non-minimal coupling STG models.
After a brief analysis of the cosmological background equations of the model,
we have studied the cosmological perturbations of the model on flat universe with three different branches.
For tensor perturbations, we have found that corrections to the friction term of GW exist on all three branches.
This would make the luminosity distance of GW to be different from the luminosity distance of the standard electromagnetic field.
We have also shown that the propagation speed of GW on one branch can be different from the speed of light,
but on the other two branches the propagation speed of GW is always equal to the speed of light.
For vector perturbations, we have demonstrated that there are two vectorial dynamic degree of freedom on one of the branches.
On this branch, the vectorial and tensorial GW have the same propagation speed and are different from the speed of light.
But there are no dynamic degree of freedom for vector perturbations on the other two branches.
This suggests the strong coupling problem on these two branches.
After examining scalar perturbations, we have proved that the model always suffers from the problem of strong coupling or ghost instability.
This conclusion also applies to the pure $f(\mathbb{Q})$ model or the pure non-minimal coupling STG model.
Since these instabilities may arise from the fact that non-trivial STG models always lack diffeomorphism (coincident gauge) or contain higher-order derivatives (via Eq.~(\ref{STGconnection})), these instabilities may exist widely in various STG models.

\subsection*{Acknowledgements}
This work is supported in part by
the National Key Research and Development Program of China under Grant No. 2020YFC2201501
and the National Natural Science Foundation of China (NSFC) under Grant No. 12147103 and 12205063.

{\it Note added}:
While this manuscript is preparing for submission, three excellent papers \cite{Gomes:2023tur,Heisenberg:2023tho,Heisenberg:2023tho2}
that also studied linear perturbations of the $f(\mbQ)$ model appeared on preprints a few days ago.
The results of this paper are consistent with their results when the model we consider is reduced back to the $f(\mbQ)$ model.


{}

\end{document}